\documentclass[prb,twocolumn,showpacs,preprintnumbers,amsmath,amssymb,10pt]{revtex4-1}

\usepackage{color}
\definecolor{blau}{rgb}{0,0,1}

\usepackage[T1]{fontenc}
\usepackage[pdftex]{graphicx}
\usepackage{dcolumn}
\usepackage{bm}

\usepackage{mathcomp}

\DeclareGraphicsExtensions{.pdf}   
\usepackage{rotating}

\DeclareMathOperator{\e}{e}

\usepackage[justification=raggedright,singlelinecheck=false]{caption}

\begin{document}

\title{Symmetry properties of vibrational modes in graphene nanoribbons}
\author{Roland Gillen}\email{rg403@cam.ac.uk}
\author{Marcel Mohr}
\author{Janina Maultzsch}
\affiliation{Institut für Festkörperphysik, Technische Universität Berlin, Hardenbergstr. 36, 10623 Berlin, Germany}

\date{\today}

\begin{abstract}
We present symmetry properties of the lattice vibrations of graphene nanoribbons with pure armchair (AGNR) and zigzag edges (ZGNR). In non-symmorphic nanoribbons the phonon modes at the edge of the Brillouin zone are twofold degenerate, whereas the phonon modes in symmorphic nanoribbons are non-degenerate. We identified the Raman-active and infrared-active modes. We predict 3$N$ and 3$(N+1)$ Raman-active modes for $N$-ZGNRs and $N$-AGNRs, respectively ($N$ is the number of dimers per unit cell). These modes can be used for the experimental characterization of graphene nanoribbons. Calculations based on density functional theory suggest that the frequency splitting of the LO and TO in AGNRs (corresponding to the $E_{2g}$ mode in graphene) exhibits characteristic width and family dependence. Further, all graphene nanoribbons have a Raman-active breathing-like mode, the frequency of which is inversely proportional to the nanoribbon width and thus might be used for experimental determination of the width of graphene nanoribbons.
\end{abstract}

\pacs{
 61.48.De 
 63.22.-m 
 63.20.D- 
 63.20.dk 
}
\maketitle

\section{Introduction}
Graphene, a single layer of carbon atoms, attracted tremendous attention for both theoretical and experimental studies. Its unique properties as a two-dimensional crystal make it a fascinating model system for fundamental studies of condensed matter physics, while its remarkable electronic and transport properties and the available fabrication by lithographic processes are promising for future application in nanoelectronic devices, $e.g.$, ballistic room-temperature transistors. However, defect-free and unpatterned graphene is a zero-gap semiconductor and is thus unsuitable for use in novel electronic devices that require precise control over carrier type and transport. 
In this context, graphene nanoribbons (GNRs), narrow stripes of graphene, have evolved from model systems for the investigation of edge effects in graphene and graphite\cite{fujita96} into promising materials in their own right. While those nanoribbons possess similarly outstanding properties as graphene, edge effects like the quantum confinement of the electronic wavefunctions in sufficiently narrow nanoribbons can result in the opening of a band gap in the electronic structure, exhibiting a characteristic dependence on chirality and nanoribbon width\cite{han206805,ezawa045432,barone606,son216803,wassmann-GNR,NikolaevGNR}. Furthermore, they give rise to interesting magnetic properties in nanoribbons with zizag-shaped edges (ZGNRs)\cite{fujita96,wakabayashi98,wakabayashi8271,kusakabe092406,yamashiro193410,sonmetall,lee174431,wassmann-GNR}. 

Lattice vibrations have a significant effect on the transport of valence electrons in graphene-related materials due to a strong electron-phonon coupling\cite{pisanibornoppen}. Therefore, the investigation of vibrational properties of graphene nanoribbons is of great interest for the physical understanding of those structures. So far, there are only a few reports addressing this topic, the most of which focus on edge-localized phonons in armchair graphene nanoribbons (AGNRs)\cite{igami-GNR, tanaka-GNR}. Kawai \emph{et al.}\cite{kawai-GNR} and Malola \emph{et al.}\cite{malola-GNR} predict the occurance of a characteristic Raman peak at a frequency about 2000 cm$^{-1}$ for nonhydrogenized armchair nanoribbons due to edge-related vibrations. Zhou \emph{et al.}\cite{zhou-GNR} calculated the lattice vibrations at the $\Gamma$-point of various AGNRs and ZGNRs and found a Raman active phonon mode of width-dependent frequency, similar to the radial breathing mode (RBM) in carbon nanotubes.
These insights could prove to be useful for characterization purposes. Yamada \emph{et al.}\cite{yamada08} used a molecular approach to derive the full phonon dispersions of small AGNRs and ZGNRs by zonefolding of the calculated phonons of polyaromatic hydrocarbons (PAHs) of suitable sizes and thus were able to relate molecular vibrations and phonons in crystals. 

However, to the best of our knowledge, no systematic work on the symmetry properties of the phonons in GNRs has been reported so far. The use of crystal symmetries is a versatile instrument in solid state physics, $e.g.$, for the theoretical investigation and understanding phonon dispersion, electronic band structure and optical activity, which are ruled by symmetry and selection rules. In this sense, not only edge effects but also the different symmetry in graphene nanoribbons compared to graphene and carbon nanotubes (CNTs) could be responsible for differences in electronic and vibrational properties. Moreover, symmetry-based selection rules for optical methods, such as Raman and IR spectroscopy, will help to predict and understand experimental results.
 
In this work, we present the symmetry properties of graphene nanoribbons with pure armchair and zizgag edges. We use group theoretical methods to derive the full dynamical representations of AGNRs and ZGNRs, describing the full symmetry properties of their phonon spectra, and discuss Raman and infrared active modes. Further, we compare our results with \textit{ab initio} density functional theory calculations of the nanoribbon phonons from previous work\cite{gillen09}.  

\section{Calculations}
The calculations of phonon frequencies of GNRs were performed with the computational package {\sc SIESTA}\cite{siesta1,siesta2}, utilizing a density functional theory approach in the local approximation form
\cite{perdew81}. Pseudopotentials were generated with the
Troullier-Martins scheme\cite{troullier91}. The valence electrons were
described by a double-$\zeta$ basis set plus an additional polarizing
orbital. The localization of the basis followed the standard split
scheme and was controlled by the energy shift, an internal {\sc SIESTA} parameter, which we set to a value of 50\,meV.
Integrations in real space were performed on a grid with
a fineness of 0.08\,\AA, which can represent plane waves up to
an energy of 270\,Ry. 30 $k$-points along the reciprocal lattice vector were used to approximate integrations in reciprocal space. The space between periodic nanoribbon images in our case was at least 20\space\AA\space in order to prevent interaction between them. We fully relaxed the atomic positions of both AGNRs and ZGNRs until atomic forces were less than 0.01 eV/\AA. The phonon calculations were performed by applying the finite difference method\cite{Yin82}. We used a supercell approach with a $9\times 9\times 1$ supercell, the above parameters for cutoff energy and basis set and a grid of 3x3x1 k-points to calculate the phonon dispersion of graphene. All frequencies were scaled by a constant factor to match the experimental frequencies of the $E_{2g}$-mode in graphene, in order to achieve a better comparability between theory and experiment.

\section{Results and Discussion}
\subsection{Symmetry Properties}
In this section, we discuss the determination of the irreducible representations that correspond to the phonons of a given nanoribbon. As graphene nanoribbons are quasi-1D-crystals and possess translational periodicity in only one direction, the description of their symmetry properties can be done using line groups\cite{vujicic-linegroups,bozovic-linegroups1,bozovic-linegroups2,linegroups}. Due to the periodicity of the crystal, all combinations of point group symmetry elements, $i.e.$, rotations, mirror planes and inversion symmetry, with translations that leave the crystal invariant, are elements of the crystal symmetry group and thus elements of the line group $L$. The action of a symmetry transformation 
\begin{equation}
R_i=(Q_i|\tau + \nu_i)\in L\label{G1}
\end{equation}
on an arbitrary vector $\mathbf{r}$ of the nanoribbon is then a transformation of $\mathbf{r}$ by the point symmetry operation $Q_i$ with a successive rigid translation by $(\tau + \nu_i)\mathbf{a}$, $i.e.$,
\begin{equation*}
(Q_i|\tau + \nu_i)\mathbf{r} = Q_i\mathbf{r} + (\tau + \nu_i)\mathbf{a}
\end{equation*}
where $\tau$ is an arbitrary integer and denotes a translation by a multiple of the lattice vector $\mathbf{a}$ and $\nu_i\in[0;1)$ denotes a translation by a fraction of $\mathbf{a}$.\\

A phonon disturbs an underlying crystal by displacing the crystal atoms by a small amount. The representation of the phonons $\Gamma^{phon}$ is thus given by the direct product of the equivalence representation $\Gamma^{eq}$, which mirrors the symmetry properties of the nanoribbon, and the representation $\Gamma^{vec}$ of an arbitrary polar vector (representing the displacement of the phonon). It can be decomposed into irreducible representations, $i.e.$,
\begin{align}
\Gamma^{phon} & = \Gamma^{eq}\otimes\Gamma^{vec}\label{eq:G2}\\
& = \sum_j{a_j\Gamma_j}\nonumber
\end{align}
where $a_j$ is the number of each irreducible representation in $\Gamma^{phon}$. Only those lattice vibrations are normal modes which are compatible with the irreducible representations of $\Gamma^{phon}$. 
The symmetry properties of the phonons with wave vector $k$ can be extracted by decomposing the equivalent representations $\Gamma^{eq}_k$ and $\Gamma^{vec}_k$ and applying Eq.~\ref{eq:G2}. The characters of the equivalence representation for the symmetry operation $R_i$ are given by 
\begin{equation}
\chi_k^{eq}(R_i)=\e^{i\mathbf{k}\cdot (\tau+\nu_i)\mathbf{a}}\sum_{i}\delta_{\{(Q_i|\tau+\nu_i)\mathbf{r}_i,\mathbf{r_i}\}}\e^{i\mathbf{K}\cdot \mathbf{r}_j}\label{eq:G3}
\end{equation}
where the delta function $\delta$ is 1 for atoms that are shifted by the symmetry operation $R_i$ from the position $\mathbf{r}_j$ to the position of an equivalent atom, and 0 otherwise. The exponential function in the sum adds a phase factor for phonons with wave vectors $\mathbf{k}$, for which $Q_i\mathbf{k}=\mathbf{k}+\mathbf{K}$, where $\mathbf{K}$ is a reciprocal lattice vector. In case of nanoribbons, $\mathbf{K}=\frac{2\pi}{a}\mathbf{e}_z$ for $k=\frac{\pi}{a}$ and $\mathbf{0}$ for all other $k$ vectors, including the $\Gamma$-point.

\subsubsection{Phonons at the $\Gamma$-point}
\begin{figure}
\centering
\includegraphics*[viewport=0 360 525 
760,width=\columnwidth]{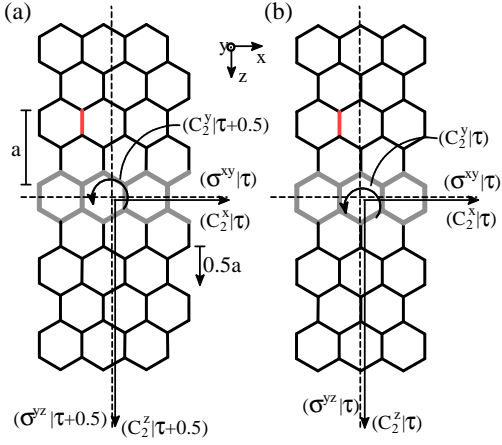}
    \caption{\label{fig:AGNRsymmetrie} (color online) Symmetry operations $(Q_i|\tau+\nu)$ for armchair nanoribbons with (a) an even number $N$ of dimers per unit cell (8-AGNR) and (b) odd number $N$ of dimers (7-AGNR). The operations $(E|t)$ and $(\sigma^{xz}|t)$, which transform each atom in the unit cell (thick grey lines) onto itself or an equivalent atom in another unit cell, are not shown. Arrows denote twofold rotations about the $x$-,$y$- and $z$-axes, dashed lines are mirror or glide planes. Dimers are emphasized by red (thin grey) lines.}
\end{figure}
\begin{figure}
\centering
\includegraphics*[viewport=0 490 525 794.97,width=\columnwidth]{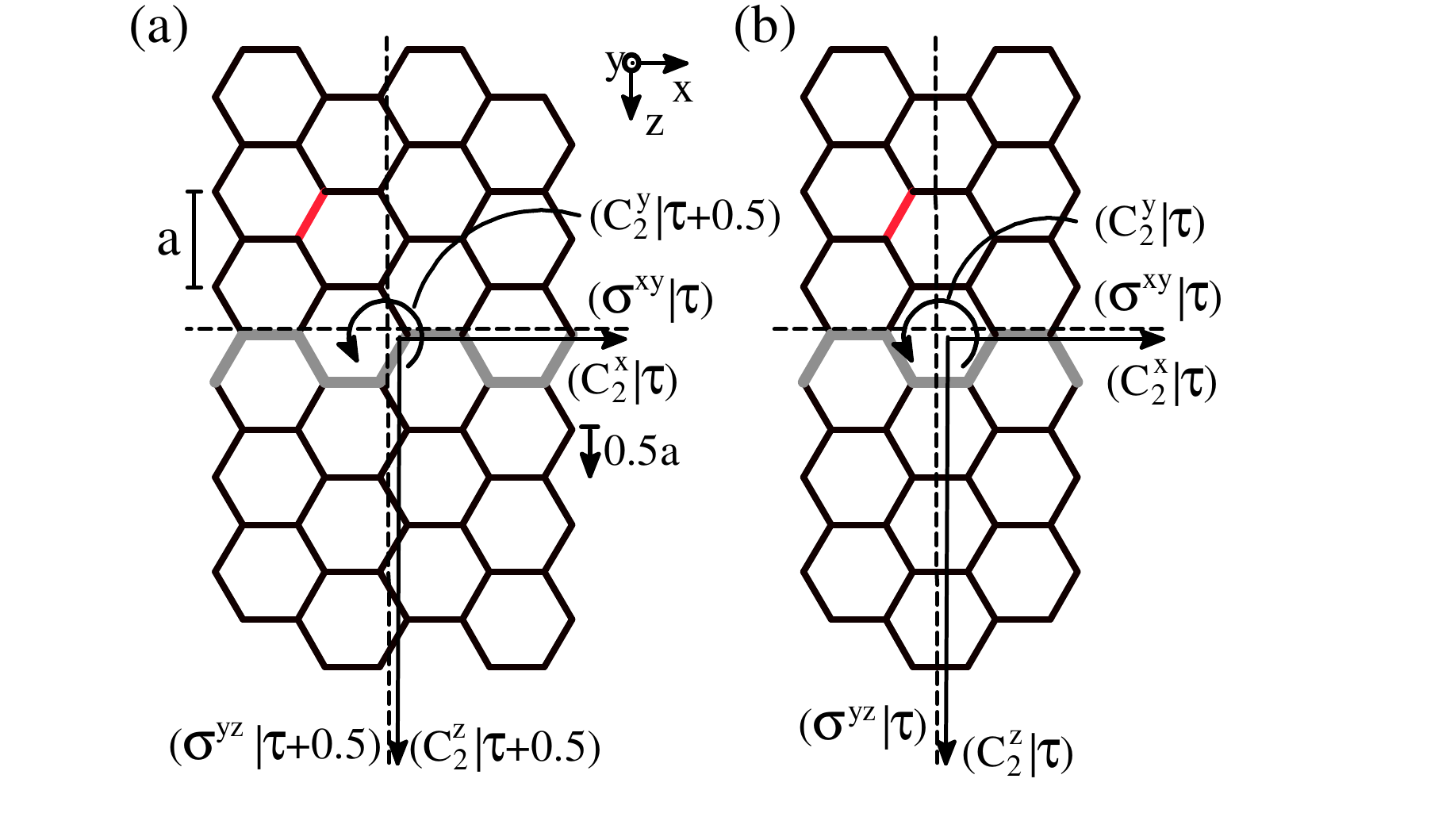}
    \caption{\label{fig:ZGNRsymmetrie} (color online) Symmetry operations $(Q_i|\tau+\nu)$ for zigzag nanoribbons with (a) an odd number $N$ of dimers per unit cell (5-ZGNR) and (b) an even number $N$ of dimers (4-ZGNR). As for Fig.~\ref{fig:AGNRsymmetrie}, $(E|\tau)$ and $(\sigma^{xz}|\tau)$ are not shown. The same notation as in Fig.~\ref{fig:AGNRsymmetrie} is used.}
\end{figure}

The symmetry properties of graphene nanoribbons 
depend on whether the number of dimers per unit cell $N$ is even or odd. As visible in Figs.~\ref{fig:AGNRsymmetrie} and~\ref{fig:ZGNRsymmetrie}, the point group components $Q_i$ of the symmetry operations $(Q_i|v_i+t)\in L$ of all nanoribbons are three two-fold rotations $C_2^x$,$C_2^y$,$C_2^z$ and three mirror operations $\sigma^{xy}$,$\sigma^{xz}$,$\sigma^{yz}$. This includes a point inversion $I=\sigma^{xy}C_2^z$. Therefore, the factor group, $i.e.$, the group of symmetry operations $R_i=(Q_i|\nu_i)$, neglecting the translations by integer multiples of the lattice vector $\mathbf{a}$, is isomorphic to the point group $D_{2h}$.

For $N$-AGNRs with odd $N$ and $N$-ZGNRs with even $N$, $v_i$ can be set to zero for all symmetry operations $R_i$ , see Fig.~\ref{fig:AGNRsymmetrie} (b) and Fig.~\ref{fig:ZGNRsymmetrie} (b). 
These nanoribbons belong to the symmorphic line group $L2/mmm$. Only atoms transformed onto equivalent atoms by the symmetry operations $(Q_i|0)$ contribute to the equivalence representation and, using the character tables for the line group $L2/mmm$ (see appendix~\ref{sec:dynrep}, Table~\ref{tab:tableL2mmm}), we obtain
\begin{align}
\Gamma^{\mbox{\tiny eq, k=0}}_{\mbox{\tiny A, odd N}}& = \frac{N+1}{2}({_{0}A_{0}^{+}} \oplus  {_{0}A_{0}^{-}}) \oplus \frac{N-1}{2} ( {_{0}A_{1}^{+}} \oplus {_{0}A_{1}^{-}})\nonumber\\
\Gamma^{\mbox{\tiny eq, k=0}}_{\mbox{\tiny Z, even N}}& = N({_{0}A_{0}^{+}} \oplus  {_{0}A_{1}^{+}})\nonumber\\
\nonumber\\
\Gamma^{\mbox{\tiny vec}}_{\mbox{\tiny k=0}} & = {_{0}A_{0}^{-}} \oplus {_{0}A_{1}^{+}} \oplus {_{0}A_{1}^{-}}\nonumber
\end{align}
where $\Gamma_A$ and $\Gamma_Z$ denote representations for AGNRs and ZGNR, respectively. The line-group notation for the irreducible representations follows Refs.~\onlinecite{bozovic-linegroups1,bozovic-linegroups2}.

The atomic displacements of corresponding atoms in different unit cells are equal for phonons with $k=0$. In this case, we can express the dynamical representations of the phonons by means of irreducible representations of the point group $D_{2h}$ (For conversion of the line-group notation into the molecular notation of irreducible representations of point groups, see appendix~\ref{sec:dynrep}, Table~\ref{tab:tableL2mmm}). Eq.~\ref{eq:G2} then yields
\begin{align}
\Gamma^{\mbox{\tiny phon, k=0}}_{\mbox{\tiny A, odd N}}& = N(A_g\oplus B_{1u}\oplus B_{2g}\oplus B_{3u})\nonumber\\ &  \oplus\frac{N+1}{2}(B_{2u}\oplus B_{3g})\nonumber\\ &  \oplus\frac{N-1}{2}(A_u\oplus B_{1g})\\
\nonumber\\
\Gamma^{\mbox{\tiny phon, k=0}}_{\mbox{\tiny Z, even N}}& = N(A_g \oplus B_{1u} \oplus B_{2g} \oplus B_{3u}\nonumber\\ &  \oplus B_{1g} \oplus B_{2u})
\end{align}

As all components for both AGNRs and ZGNRs are one-dimensional, there are no systematically degenerate phonon modes at the $\Gamma$-point.

In case of armchair nanoribbons with even $N$ and zigzag nanoribbons with odd $N$, an additional fractional translation with $v=0.5$ is needed for some symmetry operations $R_i$ to maintain crystal symmetry. These nanoribbons thus belong to the non-symmorphic line group $L2_{1}/mcm$ (appendix~\ref{sec:dynrep}, table~\ref{tab:tableL2mcm})
. In case of armchair nanoribbons, the equivalence representation changes compared to odd-$N$-AGNRs due to the fact that the nanoribbon axis does not contain any carbon atoms, see Fig.\ref{fig:AGNRsymmetrie} (a). This results in $\chi^{\mbox{\tiny eq, k=0}}_{\mbox{\tiny A, even N}}(C_2^z)=\chi^{\mbox{\tiny eq, k=0}}_{\mbox{\tiny A, even N}}(\sigma^{yz})=0$. In ZGNRs however, the altered atomic structure has no effect on $\Gamma^{\mbox{\tiny eq,k=0}}$, as the characters do not change. The equivalence representations are then
\begin{align}
\Gamma^{\mbox{\tiny eq, k=0}}_{\mbox{\tiny A, even N}}& =  N({_{0}A_{0}^{+}} \oplus  {_{0}A_{1}^{-}})\nonumber\\
\Gamma^{\mbox{\tiny eq, k=0}}_{\mbox{\tiny Z, odd N}}& = \Gamma^{\mbox{\tiny eq, k=0}}_{\mbox{\tiny Z, even N}}\nonumber\\
\nonumber\\
\Gamma^{\mbox{\tiny vec}}_{\mbox{\tiny k=0}} & = {_{0}A_{0}^{-}} \oplus {_{0}A_{1}^{+}} \oplus {_{0}A_{1}^{-}}\nonumber
\end{align}
Applying Eq.~\ref{eq:G2} results in the dynamic representation for the $\Gamma$-point phonons of $N$-AGNRs with even $N$ and ZGNRs with odd $N$:
\begin{align}
\Gamma^{\mbox{\tiny phon, k=0}}_{\mbox{\tiny A, even N}}& = N(A_g \oplus B_{1u} \oplus B_{2g} \oplus B_{3u})\nonumber\\ &  \oplus \frac{N}{2}(A_u \oplus B_{1g} \oplus B_{2u} \oplus B_{3g})\\
\nonumber\\
\Gamma^{\mbox{\tiny phon, k=0}}_{\mbox{\tiny Z, odd N}}& = N(A_g \oplus B_{1u} \oplus B_{2g} \oplus B_{3u}\nonumber\\ &  \oplus B_{1g} \oplus B_{2u})
\end{align}
Note that the equivalence representations for odd-$N$- and even-$N$-ZGNRs are equal, which implies that the symmetry properties of phonons at the $\Gamma$-point are the same for all ZGNRs.

\subsubsection{Phonons with wavevector $k$ $\neq$ $0$}
The same approach as for phonons at the $\Gamma$-point is applicable for general lattice vibrations as well, with different representations for the phonon wavevectors $\Gamma^{vec,k}$. We derived the full dynamical representations for the phonons of nanoribbons with armchair and zigzag edges, respectively, which are listed in appendix~\ref{sec:dynrep}.

The wavevectors $k$ and -$k$ are equivalent and thus representations for $k\in (0,\frac{\pi}{a})$ and $-k\in (-\frac{\pi}{a},0)$ are two-dimensional representations\cite{bozovic-linegroups2}, which have to be compatible with the representations at the $\Gamma$-point. These compatibility relations can be used to construct the representations of the phonons with 0<$|k|$<$\frac{\pi}{a}$. This is done by merging those representations of the phonons at the $\Gamma$-point that are equivalent under the lowered symmetry for 0<$|k|$<$\frac{\pi}{a}$. The displacement of the nanoribbon atoms due to a phonon with a wavevector $k$ and a frequency $\omega$ is given by
\begin{equation}
\phi(x,y,z,t) = A(x, y, x_0, y_0, z_0, t_0)\exp^{i(k(z-z_0)-\omega (t-t_0))}\label{eq:phonon}
\end{equation}
where $x_0$, $y_0$, $z_0$ are the coordinates of the origin and $t_0$ is the initial time. Eq.~\ref{eq:phonon} indicates that rotations around the $x$- and the $y$-axis [$(C_2^x|t+\nu_{Cx})$, $(C_2^y|t+\nu_{Cy})$] and the reflection through the $xy$-plane $(\sigma^{xy}|t+\nu_{\sigma xy})$ are no symmetry operations for phonons with wavevectors 0<$|k|$<$\frac{\pi}{a}$. Also, those phonons obviously possess no inversion symmetry $(I|t+\nu)$, $i.e.$, the parity of the $\Gamma$-point representations vanishes when they are extended into the Brillouin zone. As a result of the lowered symmetry and the equivalence of $k$ and -$k$, the representations $_{0}A_0^{+}$ and $_{0}A_0^{-}$, which are equivalent for the symmetry operations $(C_2^z|t+\nu_{Cz})$, $(\sigma^{xz}|t+\nu_{\sigma xz})$ and $(\sigma^{yz}|t+\nu_{\sigma yz})$, induce the two-dimensional representation $_{~~k}^{-k}E_{A_0}$. In the same way, the representations $_{0}A_1^{+}$ and $_{0}A_1^{-}$ induce the representation $_{~~k}^{-k}E_{A_1}$. $_{0}B_0^{\pm}$ and $_{0}B_1^{\pm}$ induce the representations $_{~~k}^{-k}E_{B_0}$ and $_{~~k}^{-k}E_{B_1}$, respectively (refer to Section~\ref{sec:chartab}). Thus, the dynamical representations can also be derived from compatibility relations to the $\Gamma$-point phonons instead of using Eq.~\ref{eq:G2}.

For symmorphic nanoribbons, the compatibility relations can be used for phonons at the edge of the Brillouin zone. There, the representations are one-dimensional and equivalent to the eight representations at the $\Gamma$-point. Pairs of the two-dimensional representations for $k=(0,\frac{\pi}{a})$, $_{~~k}^{-k}E_{X_y}$, transform into one dimensional representations of the same type, one representation of even parity ($X_y^+$) and one with odd parity ($X_y^-$), respectively. Thus, there is no systematic degeneracy of phonon modes at the edge of the Brillouin zone for symmorphic nanoribbons.

For non-symmorphic nanoribbons, pairs of representations ${_{~~k}^{-k}E_{A_0}}$ and ${_{~~k}^{-k}E_{A_1}}$ (${_{~~k}^{-k}E_{B_0}}$ and ${_{~~k}^{-k}E_{B_1}}$) induce two-dimensional representations  ${_{\frac{\pi}{a}}E_{A_0}^{A_1}}$ (${_{\frac{\pi}{a}}E_{B_0}^{B_1}}$) at $k=\frac{\pi}{a}$. Consequently, in non-symmorphic nanoribbons, all phonon modes are twofold degenerate at the edge of the Brillouin zone.\\
The symmetry-induced degeneracy of the phonon modes at the Brillouin zone can be clearly seen in \textit{ab initio} calculations of the full phonon dispersions of small-width nanoribbons presented in previous work\cite{gillen09}. We found that almost all branches of the phonon dispersion of AGNRs with even $N$ and ZGNRs with odd $N$, $i.e.$, the non-symmorphic nanoribbons, are twofold degenerate at $k=\frac{\pi}{a}$. No systematic degeneracy was found for the symmorphic nanoribbons, $i.e.$, $N$-AGNRs with odd $N$. Corresponding results have been reported by Yamada et al\cite{yamada08}. These results were not understood from simple zonefolding considerations. Here we have shown that instead the degeneracy of phonon modes at the Brillouin zone edge is solely determinded by the symmetry properties of the nanoribbons. We refer to Fig.~9 in Ref.~[\onlinecite{gillen09}] and Figs.~3 and~9 in Ref.~[\onlinecite{yamada08}] for calculated phonon dispersions of various AGNRs and ZGNRs.

\subsection{Raman active modes}
In this section, we will analyse which of the nanoribbon phonon modes can be observed in a Raman experiment. Raman scattering is inelastic scattering of light, where a change in frequency of the scattered light occurs due to the creation or the annihilation of Raman active phonons in the material. The polarization and the intensity of the scattered light are linked to the incident light by the Raman tensor $\mathcal{R}$; the scattering intensity is proportional to
\begin{equation*}
|\mathbf{e}_s\cdot\mathcal{R}\cdot\mathbf{e}_i|^2
\end{equation*}
where $\mathbf{e}_i$ and $\mathbf{e}_s$ denote the polarization of incoming and scattered light, respectively. As the Raman tensor is a second-rank tensor, its symmetry $\Gamma^{\mbox{\tiny Raman}}$ is given by the direct product of the representations of a polar vector, $\Gamma^{vec, k=0}$ and $\Gamma^{vec, k=0^*}$, $i.e.$,
\begin{align}
\Gamma^{\mbox{\tiny Raman}} & = \Gamma^{vec,k=0}\otimes \Gamma^{vec,k=0^*}\nonumber\\
& = A_g \oplus B_{1g} \oplus B_{2g} \oplus B_{3g}
\end{align}
Only the phonon components contained in the representation of the Raman tensor are Raman active. All Raman active phonon modes are even (subscript $g$) under inversion.\\
The polarization of the scattered light in relation to the polarization of the incident light depends on the involved Raman active phonon, see Table~\ref{tab:polarization}. In the case that a phonon with $A_g$ symmetry is emitted or absorbed, the polarization of the scattered light is parallel to that of the incident light. In case of $B_{1g}$, incoming and scattered light are cross-polarized in the $xy$-plane. Similarly, $B_{2g}$ and $B_{3g}$ change the polarization of the incoming light in the $xz$-plane and $yz$-plane, respectively.
\begin{table}[tb]
\caption{\label{tab:polarization} Allowed Raman scattering configurations.}
\begin{tabular}{c|c}
phonon symmetry&allowed scattering configuration ($\mathbf{e}_i$,  $\mathbf{e}_s$)\\
\hline
$A_g$&($z$,$z$)\\
$B_{1g}$&($x$,$y$)\\
$B_{2g}$&($x$,$z$)\\
$B_{3g}$&($y$,$z$)\\
\end{tabular} 
\end{table}

In a former work, we showed that it is possible to classify the $\Gamma$-point phonon modes of nanoribbons into six fundamental modes and $6(N-1)$ overtones, $N-1$ overtones for each fundamental mode. The six fundamental modes are similar to the six phonon branches at the $\Gamma$-point of graphene, thus they are of special interest. The symmetry properties of the acoustic fundamental modes are equal for all AGNRs and ZGNRs, and none of these modes is Raman active.  The longitudinal-acoustic phonon (0-LA), which induces an atomic displacement parallel to the nanoribbon axis, has $B_{1u}$ symmetry. The representations of the transverse-acoustic (0-TA) and the out-of-plane-acoustic fundamental mode (0-ZA) are $B_{3u}$ and $B_{2u}$, respectively, for all nanoribbons.
In contrast, the representations of the three optical fundamental modes are of even parity, and consequently, these phonons are Raman active. For AGNRs, the longitudinal-optical fundamental mode (0-LO) has $A_g$ symmetry and can thus be observed only for $zz$-polarization. The transverse-optical (0-TO) and out-of-plane-optical modes (0-ZO) have $B_{2g}$ and $B_{3g}$ symmetry, respectively, and are forbidden in $zz$-polarizaton. We therefore expect that the three optical modes can be observed separately by polarization-dependent Raman measurements.\\
For ZGNRs, the representations of the 0-LO and 0-TO are interchanged compared to AGNRs ($i.e.$, 0-LO: $B_{2g}$, 0-TO: $A_g$) and the 0-ZO has $B_{1g}$ symmetry.
The reason for this is the difference in structure between AGNRs and ZGNRs. In ZGNRs, the twofold rotation $C_2^{x}$ around the $x$-axis of the nanoribbon and the mirror plane $\sigma{xy}$ leave carbon atoms invariant, in constrast to AGNRs. In this sense, there is a simple correspondence between ZGNRs and armchair nanotubes, where the TO is fully symmetric and Raman-allowed in $zz$-configuration, and between AGNRs and zigzag nanotubes, where the LO phonon is fully symmetric. Note that in AGNRs and ZGNRs both LO and TO are Raman active, though in different scattering geometries, whereas in zigzag and armchair carbon nanotubes the TO and LO, respectively, are Raman inactive\cite{reich04buch}. For characterization of nanoribbons or determination of the graphene crystallographic direction, the different LO and TO selection rules in AGNRs and ZGNRs might be useful.

\begin{figure}
\centering
\includegraphics*[width=\columnwidth]{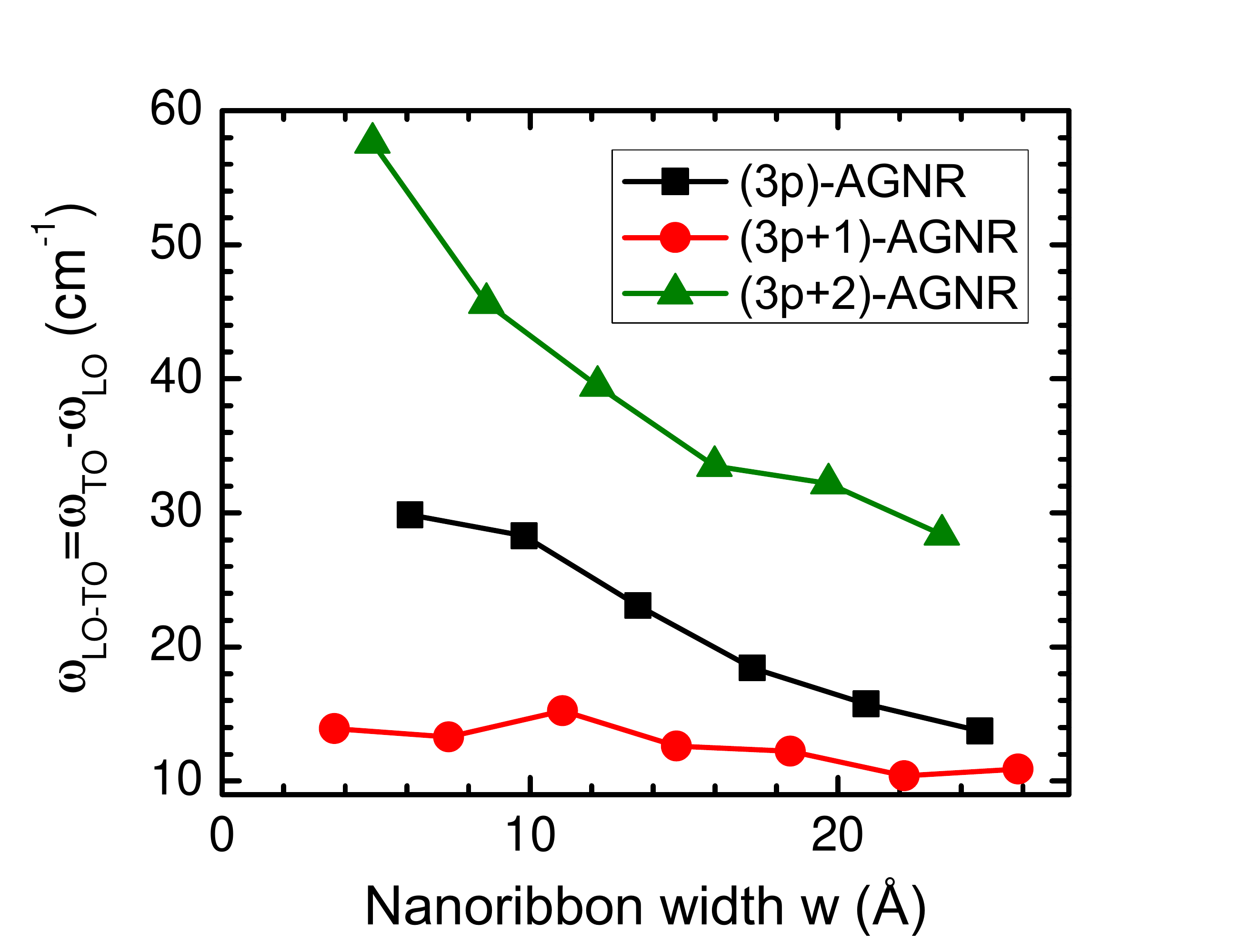}
\caption{\label{fig:TOLO} (color online) 
Difference in frequency between the transverse-optical (0-TO) and the longitudinal\--optical (0-LO) fundamental modes in armchair nanoribbons. The size of the LO-TO-splitting is noticeable different for AGNRs of different families.
}
\end{figure}

As shown in Ref.~\onlinecite{gillen09}, there are three families of armchair nanoribbons regarding the $\Gamma$-point frequencies of the 0-LO and the 0-TO. This corresponds to the behaviour of the electronic properties\cite{son216803}, where AGNRs were found to be quasi-metallic or semiconducting, depending on the number of dimers in the unit cell. The magnitude of the band gap can be classified into distinct families. This family behaviour is also found for the LO-TO-splitting, see Fig.~\ref{fig:TOLO}
. The quasi-metallic nanoribbons of the ($N$=3p+2)-family possess the largest splitting of the 0-LO and the 0-TO, which indicates a Kohn-anomaly at the $\Gamma$-point for these nanoribbons. The frequency splitting quickly decreases with increasing nanoribbon width $w$.
In comparison, the nanoribbons of the ($N$=3p)-family, which are semiconducting with intermediately sized band gaps, exhibit a similar but weaker drop in frequency splitting. (3p+1)-AGNRs, possessing the largest band gaps, exhibit only a weak dependence of the LO-TO-splitting on the nanoribbon width. We found an overall small decrease from 14 cm$^{-1}$ for a 4-AGNR to 11 cm$^{-1}$ for a 22-AGNR.

\subsubsection{Raman activity of overtones}
\begin{table}
\caption{\label{tab:ramanover} Symmetries of the Raman active overtones in AGNRs and ZGNRs. Due to inversion symmetry, the Raman active overtones of optical fundamental modes are all of even vibrational order. Similarly, the Raman active 'acoustic' overtones possess an odd vibrational order.}
\begin{ruledtabular}
\begin{tabular}{ccccccc}
 &\multicolumn{3}{c}{$n$ even}&\multicolumn{3}{c}{$n$ odd}\\
 &$n$-LO&$n$-TO&$n$-ZO&$n$-LA&$n$-TA&$n$-ZA\\ \hline
 AGNR&$A_g$&$B_{2g}$&$B_{3g}$&$B_{2g}$&$A_g$&$B_{1g}$\\\hline
 ZGNR&$B_{2g}$&$A_g$&$B_{1g}$&$B_{2g}$&$A_g$&$B_{1g}$\\
\end{tabular}
\end{ruledtabular}
\end{table}
An inspection of the symmetry properties of overtones
shows that all overtones with an even vibrational order possess the same symmetry properties as their respective fundamental modes (at the $\Gamma$-point), $e.g.$, 0-LO, 2-LO, 4-LO, etc. in an AGNR are all of $A_g$ symmetry. All these modes have in common that there is no vibrational "node" coinciding with the nanoribbon axis, see discussion in Ref.~\onlinecite{gillen09}.\\
In the same way, the overtones with an odd vibrational order of a particular fundamental mode possess the same irreducible representation, and there is always a vibrational node on the $xz$-plane of the nanoribbon. A major point regarding the Raman activity is the inversion symmetry of the phonons. The acoustic fundamental modes and their overtones of even order are antisymmetric with respect to inversion and are thus not Raman active. The vibrational node on the nanoribbon axis in case of odd-order overtones of the acoustic modes, however, induces a inversion symmetry. Consequenty, those phonons are Raman active, see Table~\ref{tab:ramanover}. For the optical phonons, overtones of even vibrational order possess even parity under inversion. However, the overtones with odd vibrational order possess odd parity under inversion and consequently are not Raman active.\\
A nanoribbon with an even number $N$ of dimers per unit cell has $\frac{N}{2}-1$ overtones of even vibrational order and $\frac{N}{2}$ overtones of odd vibrational order for each fundamental mode. In case of a nanoribbon with an odd number of dimers, there are $\frac{N-1}{2}$ overtones of even and $\frac{N-1}{2}$ overtones of odd vibational order. 
Thus, we would expect to find 3$N$ Raman active phonons in case of nanoribbons with even $N$ and 3$(N+1)$ Raman active phonons in case of nanoribbons with an odd number $N$ of dimers per unit cell.
The Raman active modes and their symmetries are summarized in Table~\ref{tab:ramanover}.

Regarding spectroscopic characterization we note that the $B_{3g}$ symmetry is found only for AGNRs for the out-of-plane-optical fundamental mode and its  overtones. This could provide a way to experimentally distingiush armchair and zigzag edges in nanoribbons by cross-polarized Raman measurements, see Tables~\ref{tab:polarization} and~\ref{tab:ramanover}. However, it will be experimentally difficult to perform a measurement with light polarization perpendicular to the nanoribbon plane.

\subsubsection{Breathing-like mode in nanoribbons} 
\begin{figure}
\centering
\begin{minipage}[b]{0.95\columnwidth}
\includegraphics[width=\columnwidth]{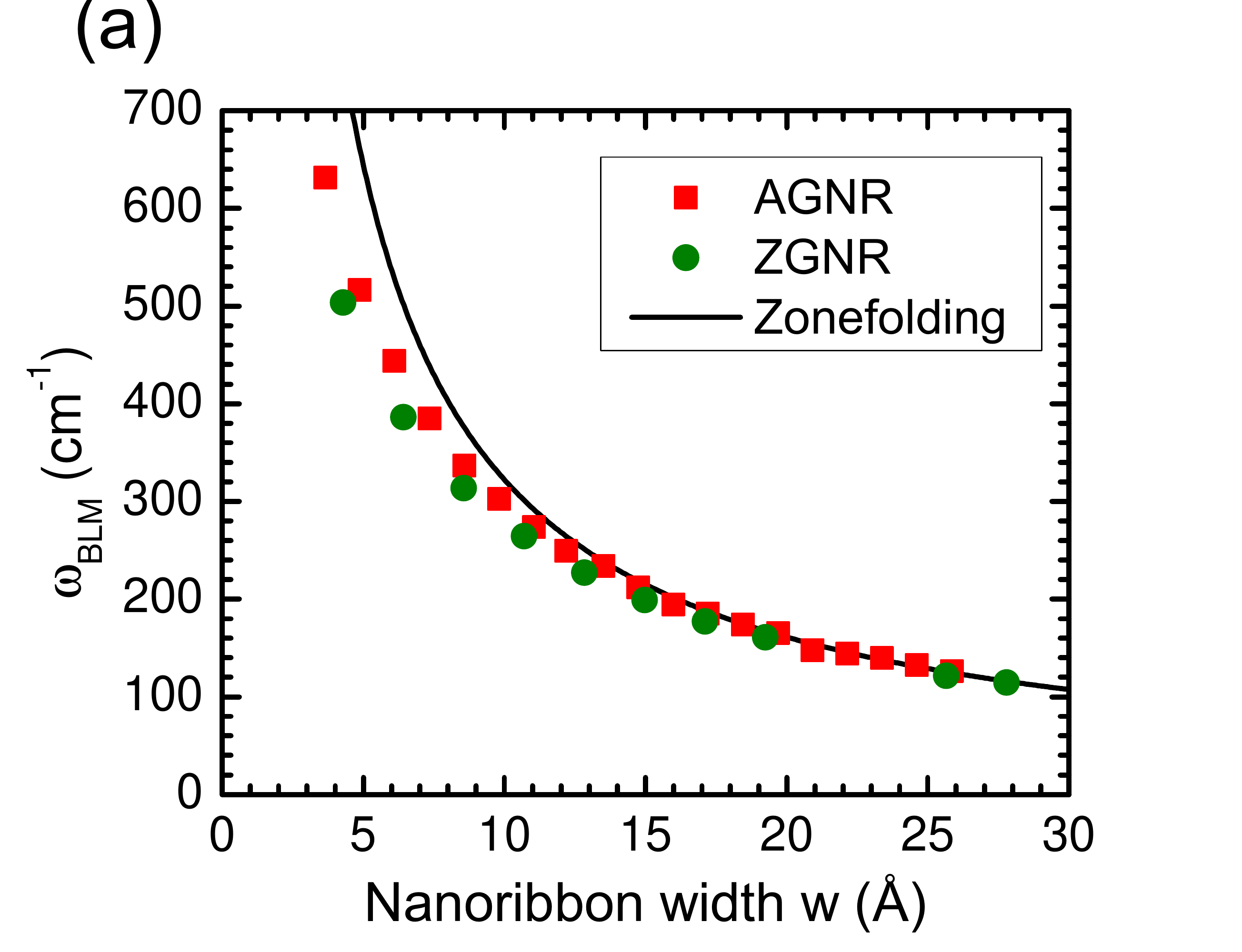}
\end{minipage}
\\
\begin{minipage}[b]{0.95\columnwidth}
\includegraphics[width=0.6\columnwidth]{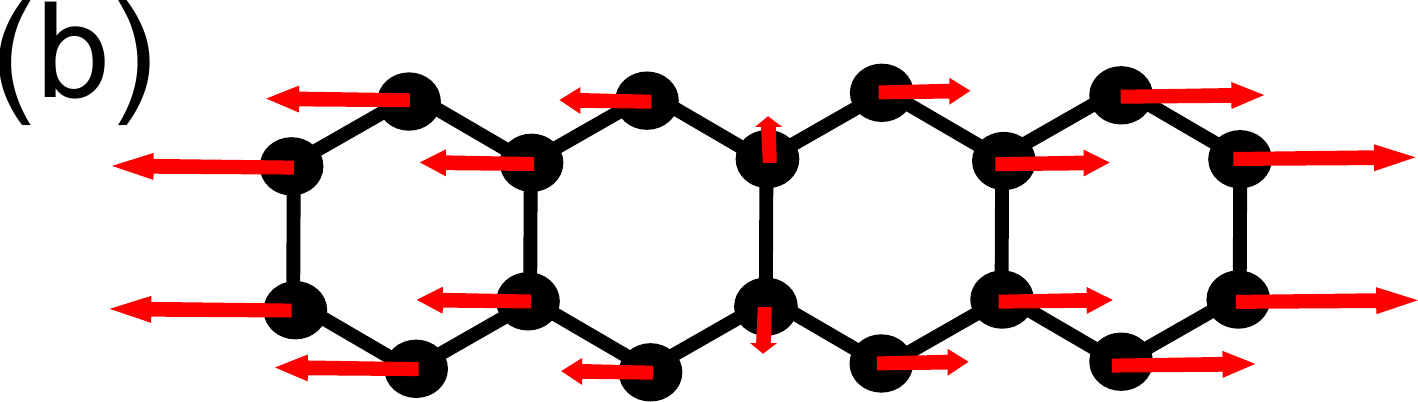}
\end{minipage}
\caption{\label{fig:BLM} (Color online)
Frequency of the breathing-like mode (BLM) in armchair (red squares) and zigzag (green circles) nanoribbons of different widths $w$. Zonefolding of the linear part of the LA branch in graphene results in an approximation for the BLM of $\omega_{\textrm{BLM}}= \frac{a\dot\pi}{w}$, where a=1025.6 \AA$\cdot$ cm$^{-1}$ is the sound velocity in graphene.ction of the BLM phonon. The nanoribbon is uniformly expanded and compressed.
}
\end{figure}

Another promising phonon mode for characterization purposes is the first overtone of the transverse-acoustic fundamental mode (1-TA). Here, all atoms of one half of the nanoribbon move in-phase and in opposite direction than the atoms in the other half, as shown in Fig.~\ref{fig:BLM}~(b). This results in a breathing-like expansion and compression of the nanoribbon and shows great similarities to the radial breathing mode (RBM) of carbon nanotubes. The displacement pattern of this phonon mode is symmetric under inversion, thus it is Raman active. The symmetry of the mode is $A_g$ (refer to Table~\ref{tab:ramanover}). From a theoretical point of view, its frequency should exhibit a characteristic dependence of the nanoribbon width.
Fig.~\ref{fig:BLM}~(a) shows the calculated frequencies of BLMs of AGNRs and ZGNRs of various widths. The frequencyies of the breathing-like modes display a strong dependence on the nanoribbon width and is nearly independent of the edge type of the nanoribbon, $i.e.$, the BLM has almost the same frequency for AGNRs and ZGNRs of equal widths. Small deviations are found for the smallest nanoribbons.

\begin{figure}
\centering
\includegraphics[width=\columnwidth]{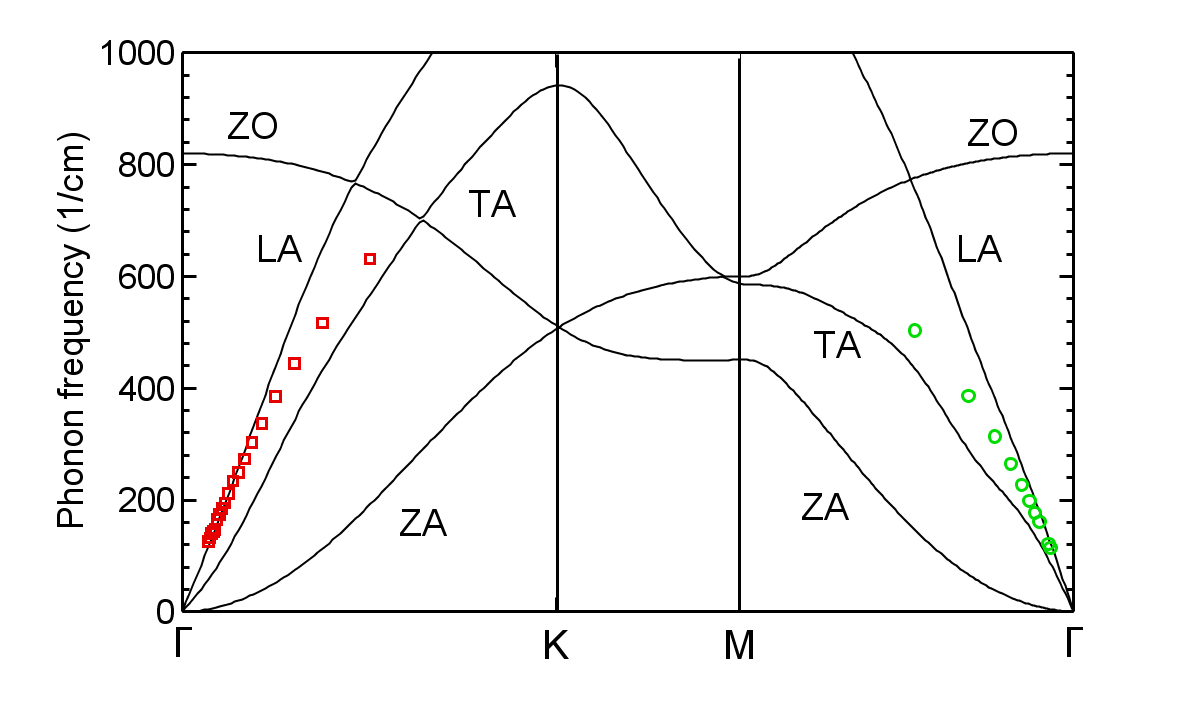}
    \caption{\label{fig:BLM-disp} (Color online) Mapping of the BLM frequencies of nanoribbons of various widths onto the phonon dispersion of graphene. As the BLM is the first overtone of the nanoribbon TA fundamental mode, the corresponding wave vector decreases with increasing nanoribbon width. Due to different orientation of the nanoribbon and the graphene lattice, the nanoribbon TA corresponds to the graphene LA. The BLM generally lies on the graphene LA branch, except for nanoribbons of very small widths (=larger wave vectors). The calculated nanoribbon frequencies have been rescaled by the same constant factor as the graphene frequencies.}
\end{figure}

We assigned wavelengths and wave vectors to the BLMs of various nanoribbons and mapped them onto the phonon dispersion of graphene. We showed in a previous work that the $\Gamma$-point phonons of AGNRs can be unfolded to reproduce the $\Gamma$KM direction in graphene, while vibrations of ZGNRs reproduce the phonon dispersion in $\Gamma$M direction, respectively. In this way, the BLMs of nanoribbons of increasing width should move along the LA branch of graphene towards the $\Gamma$-point in the phonon dispersion. Figure~\ref{fig:BLM-disp} shows such a mapping of the BLMs onto the phonon dispersion of graphene. It can be seen that the BLMs of AGNRs and ZGNRs exhibit deviations from the LA branch for smaller nanoribbons, but comply very well in the case of larger nanoribbons. 
The agreement of our calculated BLM frequencies with the linear part of the graphene LA branch enables us to derive a formula for estimation of the BLM frequency of larger nanoribbons. In the linear part of the branch, the dispersion is given by
\begin{equation*}
\omega = a\cdot\frac{2\pi}{\lambda}
\end{equation*}

where $k$ is the wave length of the vibration and $a = 1025.6$ \AA$\cdot$ cm$^{-1}$ is the sound velocity in our calulated phonon dispersion of graphene. The wave length of the BLM is $\lambda = 2\dot w$ for all nanoribbons, where $w$ is the nanoribbon width. The frequency of the BLM of not too narrow nanoribbons can then be estimated by
\begin{eqnarray}
\omega_{BLM} &=& \frac{a\cdot\pi}{w}\label{eq:zonefolding}\\
&=& \frac{3222\textrm{\AA$\cdot$ cm$^{-1}$}}{w}\nonumber
\end{eqnarray}

In this zonefolding model, the BLM of a nanoribbon with a width of 10 nm would then have a frequency of approximately 32 cm$^{-1}$. We expect that the agreement of the BLM frequencies with the LA branch of graphene further improves for nanoribbons of increasng width. It is thus likely that the Eq.~\ref{eq:zonefolding} can be used for estimating the BLM frequencies of nanoribbons with realistic widths $w$ > 10 nm.
 
\subsection{Infrared activity}
Another spectroscopic method for determining vibrational properties is the exitation of phonons by absorbtion of infrared light. For a phonon to be infrared active, it has to change the permanent dipole moment in the solid and thus be able to couple with an electromagnetic field. The representation $\Gamma^{\mbox{\tiny IR}}$ of IR active phonons thus transforms like the representation of a polar vector $\Gamma^{vec}$, $i.e.$,

\begin{align}
\Gamma^{\mbox{\tiny IR}} & = \Gamma^{vec}\nonumber\\
& = B_{1u}\oplus B_{2u}\oplus B_{3u}
\end{align}
The $u$ indicates that the decomposition of $\Gamma^{\mbox{\tiny IR}}$ consists of only "odd" irreducible representations. Consequently, for nanoribbons, all $\Gamma$-point phonons that are not Raman active are IR active, $i.e.$, the three acoustic fundamental modes and the respective overtones of even vibrational order and the odd overtones of optical fundamental modes. Thus, we would expect to find 3$N$ IR active phonons for all nanoribbons, see Table~\ref{tab:irover}.
\begin{table}[bt]
\caption{\label{tab:irover} Symmetries of the infrared (IR) active overtones in AGNRs and ZGNRs. As all nanoribbons possess inversion symmetry, only phonons that are not Raman active are infrared active.}
\begin{ruledtabular}
\begin{tabular}{ccccccc}
 &\multicolumn{3}{c}{$n$ odd}&\multicolumn{3}{c}{$n$ even}\\
 &$n$-LO&$n$-TO&$n$-ZO&$n$-LA&$n$-TA&$n$-ZA\\ \hline
 AGNR&$B_{3u}$&$B_{1u}$&$A_u$&$A_u$&$B_{3u}$&$B_{2u}$\\\hline
 ZGNR&$A_u$&$B_{3u}$&$B_{2u}$&$A_u$&$B_{3u}$&$B_{2u}$\\
\end{tabular}
\end{ruledtabular}
\end{table}

\section{Conclusion}
We used group theory to study the symmetry properties of graphene nanoribbons with pure armchair- and zigzag-type edges and derived the full dynamical representations for these nanoribbons. The different symmetry properties of the phonons of nanoribbons with symmorphic and non-symmorphic symmetry groups mainly manifest themselfs at the edge of the Brillouin zone. Here, all phonon modes of non-symmorphic nanoribbons are two-fold degenerate, whereas the phonon modes of symmorphic nanoribbons are always non-degenerate. These theoretical derivations explain results from previous \textit{ab initio} calculations\cite{gillen09}, where we calculated the full phonon dispersions of small-width AGNRs and ZGNRs. They also explain the differences in degeneracy at the edge of the Brillouin zone between graphene nanoribbons and carbon nanotubes. Using the dynamical representations at the $\Gamma$-point, we predict the phonon spectra of graphene nanoribbons to consist of 3$N$ Raman active and 3$N$ infrared active vibrational modes.
Some Raman active modes are promising candidates for experimental characterization of nanoribbons. Our calculations of phonon frequencies using density functional theory suggest that the frequency splitting of the longitudinal-optical and the transverse-optical fundamental mode (corresponding to the TO and the LO of graphene at the $\Gamma$-point) exhibits a characteristic family dependence, with quasi-metallic nanoribbons possessing the largest splittings. Further, we found a Raman active breathing-like mode (BLM) with a strong width-dependence, similar to the radial breathing mode in carbon nanotubes. Our results might prove to be useful for future experimental investigations of the vibrational properties of graphene nanoribbons.

\section{Acknowledgements}
We thank C. Thomsen for his support and helpful discussions. 
This work was supported in part by the Cluster of
Excellence 'Unifying Concepts in Catalysis' coordinated by the TU Berlin and
funded by DFG.

\appendix
\begin{widetext}
\section{Character tables}\label{sec:chartab}
Here, we show the character tables for the irreducible representations of the two symmetry groups that were relevant for this paper, the symmorphic $L2/mmm$ group and the non-symmorphic $L2/mcm$ group. Both character tables were adapted from previous work by Bozovic \emph{et al.}\cite{bozovic-linegroups2}. The corresponding matrices can be found in the same work.
\begin{table*}[b]
\caption{\label{tab:tableL2mmm} Characters of the irreducible representations of the symmorphic group $L2/mmm$ for the k-points $k=[0;\frac{\pi}{a}]$. For $0<k<\frac{\pi}{a}$, representations for $k$ and $-k$ generate two-dimensional representations ${^{-k}_{~~k}E_{X}}$, whereas the representations for $k=0$ and $k=\frac{\pi}{a}$ are one-dimensional. $\delta_0=2\cos(k\tau a)$. The first 
column shows the relevant basis functions for each $k$ for the purposes of this paper. The irreducible representations of the symmetry group $L2/mmm$ for different k-points appear in two different notations in the second column. On the left side is the line group notation (LGN), on the right side is the molecular notation (MN), which is only applicable for phonons at the $\Gamma$-point. 
}
\begin{tabular}{r|cc|cccccccc}   Basis&LGN&MN&($E$|$\tau$)&($C_2^x$|$\tau$)&($C_2^y$|$\tau$)&($C_2^z$|$\tau$)&(i|$\tau$)&($\sigma^{yz}$|$\tau$)&($\sigma^{xz}$|$\tau$)&($\sigma^{xy}$|$\tau$)\\
\hline
&${_0A_0^+}$&$A_g$&1&1&1&1&1&1&1&1\\
z&${_0A_0^-}$&$B_{1u}$&1&-1&-1&1&-1&1&1&-1\\
&${_0B_0^+}$&$B_{1g}$&1&-1&-1&1&1&-1&-1&1\\
&${_0B_0^-}$&$A_u$&1&1&1&1&-1&-1&-1&-1\\
x&${_0A_1^+}$&$B_{3u}$&1&1&-1&-1&-1&-1&1&1\\
&${_0A_1^-}$&$B_{2g}$&1&-1&1&-1&1&-1&1&-1\\
y&${_0B_1^+}$&$B_{2u}$&1&-1&1&-1&-1&1&-1&1\\
&${_0B_1^-}$&$B_{3g}$&1&1&-1&-1&1&1&-1&-1\\
z&${^{-k}_{~~k}E_{A_0}}$&&$\delta_0(\tau)$&0&0&$\delta_0(\tau)$&0&$\delta_0(\tau)$&$\delta_0(\tau)$&0\\
&${^{-k}_{~~k}E_{B_0}}$&&$\delta_0(\tau)$&0&0&$\delta_0(\tau)$&0&-$\delta_0(\tau)$&-$\delta_0(\tau)$&0\\
x&${^{-k}_{~~k}E_{A_1}}$&&$\delta_0(\tau)$&0&0&-$\delta_0(\tau)$&0&-$\delta_0(\tau)$&$\delta_0(\tau)$&0\\
y&${^{-k}_{~~k}E_{B_1}}$&&$\delta_0(\tau)$&0&0&-$\delta_0(\tau)$&0&$\delta_0(\tau)$&-$\delta_0(\tau)$&0\\
&${_{\frac{\pi}{a}} A_0^+}$&&(-1)$^{\tau}$&(-1)$^{\tau}$&(-1)$^{\tau}$&(-1)$^{\tau}$&(-1)$^{\tau}$&(-1)$^{\tau}$&(-1)$^{\tau}$&(-1)$^{\tau}$\\
z&${_{\frac{\pi}{a}} A_0^-}$&&(-1)$^{\tau}$&-(-1)$^{\tau}$&-(-1)$^{\tau}$&(-1)$^{\tau}$&-(-1)$^{\tau}$&(-1)$^{\tau}$&(-1)$^{\tau}$&-(-1)$^{\tau}$\\
&${_{\frac{\pi}{a}} B_0^+}$&&(-1)$^{\tau}$&-(-1)$^{\tau}$&-(-1)$^{\tau}$&(-1)$^{\tau}$&(-1)$^{\tau}$&-(-1)$^{\tau}$&-(-1)$^{\tau}$&(-1)$^{\tau}$\\
&${_{\frac{\pi}{a}}B_0^-}$&&(-1)$^{\tau}$&(-1)$^{\tau}$&(-1)$^{\tau}$&(-1)$^{\tau}$&-(-1)$^{\tau}$&-(-1)$^{\tau}$&-(-1)$^{\tau}$&-(-1)$^{\tau}$\\
x&${_{\frac{\pi}{a}}A_1^+}$&&(-1)$^{\tau}$&(-1)$^{\tau}$&-(-1)$^{\tau}$&-(-1)$^{\tau}$&-(-1)$^{\tau}$&-(-1)$^{\tau}$&(-1)$^{\tau}$&(-1)$^{\tau}$\\
&${_{\frac{\pi}{a}} A_1^-}$&&(-1)$^{\tau}$&-(-1)$^{\tau}$&(-1)$^{\tau}$&-(-1)$^{\tau}$&(-1)$^{\tau}$&-(-1)$^{\tau}$&(-1)$^{\tau}$&-(-1)$^{\tau}$\\
y&${_{\frac{\pi}{a}} B_1^+}$&&(-1)$^{\tau}$&-(-1)$^{\tau}$&(-1)$^{\tau}$&-(-1)$^{\tau}$&-(-1)$^{\tau}$&(-1)$^{\tau}$&-(-1)$^{\tau}$&(-1)$^{\tau}$\\
&${_{\frac{\pi}{a}} B_1^-}$&&(-1)$^{\tau}$&(-1)$^{\tau}$&-(-1)$^{\tau}$&-(-1)$^{\tau}$&(-1)$^{\tau}$&(-1)$^{\tau}$&-(-1)$^{\tau}$&-(-1)$^{\tau}$\\
\end{tabular} 
\end{table*}

\clearpage

\begin{table*}[t]
\caption{\label{tab:tableL2mcm} Characters of the irreducible representations of the non-symmorphic group $L2_1/mcm$ for the k-points $k=(0;\frac{\pi}{a}]$. As for the symmorphic counterpart (table~\ref{tab:tableL2mmm}), the representations for $0<k<\frac{\pi}{a}$ are two-dimensional, so are the representations for $k=\frac{\pi}{a}$, which are generated by ${^{-k}_{~~k}E_{A_0}}$,${^{-k}_{~~k}E_{A_1}}$ and ${^{-k}_{~~k}E_{B_0}}$,${^{-k}_{~~k}E_{B_1}}$\cite{bozovic-linegroups2}. $\delta_0=2\cos(k\tau a)$ and $\delta_{\frac{1}{2}}=2\cos(k(\tau +\frac{1}{2})a)$. 
}
\begin{tabular}{r|cc|cccccccc}
Basis&LGN&MN&($E$|$\tau$)&($C_2^x$|$\tau$)&($C_2^y$|$\tau+\frac{1}{2}$)&($C_2^z$|$\tau+\frac{1}{2}$)&(i|$\tau+\frac{1}{2}$)&($\sigma^{yz}$|$\tau+\frac{1}{2}$)&($\sigma^{xz}$|$\tau$)&($\sigma^{xy}$|$\tau$)\\
\hline
&${_0A_0^+}$&$A_g$&1&1&1&1&1&1&1&1\\
z&${_0A_0^-}$&$B_{1u}$&1&-1&-1&1&-1&1&1&-1\\
&${_0B_0^+}$&$B_{1g}$&1&-1&-1&1&1&-1&-1&1\\
&${_0B_0^-}$&$A_u$&1&1&1&1&-1&-1&-1&-1\\
x&${_0A_1^+}$&$B_{3u}$&1&1&-1&-1&-1&-1&1&1\\
&${_0A_1^-}$&$B_{2g}$&1&-1&1&-1&1&-1&1&-1\\
y&${_0B_1^+}$&$B_{2u}$&1&-1&1&-1&-1&1&-1&1\\
&${_0B_1^-}$&$B_{3g}$&1&1&-1&-1&1&1&-1&-1\\
z&${^{-k}_{~~k}E_{A_0}}$&&$\delta_0(\tau)$&0&0&$\delta_{\frac{1}{2}}(\tau)$&0&$\delta_{\frac{1}{2}}(\tau)$&$\delta_0(\tau)$&0\\
&${^{-k}_{~~k}E_{B_0}}$&&$\delta_0(\tau)$&0&0&$\delta_{\frac{1}{2}}(\tau)$&0&-$\delta_{\frac{1}{2}}(\tau)$&-$\delta_0(\tau)$&0\\
x&${^{-k}_{~~k}E_{A_1}}$&&$\delta_0(\tau)$&0&0&-$\delta_{\frac{1}{2}}(\tau)$&0&-$\delta_{\frac{1}{2}}(\tau)$&$\delta_0(\tau)$&0\\
y&${^{-k}_{~~k}E_{B_1}}$&&$\delta_0(\tau)$&0&0&$\delta_{\frac{1}{2}}(\tau)$&0&$\delta_{\frac{1}{2}}(\tau)$&$\delta_0(\tau)$&0\\
&${_{\frac{\pi}{a}}E_{A_0}^{A_1}}$&&2(-1)$^{\tau}$&0&0&0&0&0&2(-1)$^{\tau}$&0\\
&${_{\frac{\pi}{a}}E_{B_0}^{B_1}}$&&2(-1)$^{\tau}$&0&0&0&0&0&-2(-1)$^{\tau}$&0\\
\end{tabular}
\end{table*}

\section{Dynamical representations of the phonons in armchair- and zigzag-edged graphene nanoribbons}\label{sec:dynrep}
Here, we report the full dynamical representations of the phonons in ideal armchair- and zigzag-edged nanoribbons, as follows by Eq.~\ref{eq:G2}.
\begin{align}
\Gamma^{\mbox{\tiny phon}}_{\mbox{\tiny A, odd}} = &
N( {_{0}A_{0}^{+}} \oplus {_{0}A_{0}^{-}} \oplus {_{0}A_{1}^{-}} \oplus {_{0}A_{1}^{+}} ) \oplus \frac{N+1}{2}( {_{0}B_{1}^{+}} \oplus {_{0}B_{1}^{-}}) \oplus \frac{N-1}{2}( {_{0}B_{0}^{-}} \oplus {_{0}B_{0}^{+}} )\nonumber\\ 
& \quad \oplus \sum_k\{2N({_{~~k}^{-k}E_{A_0}} \oplus 2{_{~~k}^{-k}E_{A_1}}) \oplus (N-1){_{~~k}^{-k}E_{B_0}} \oplus (N+1){_{~~k}^{-k}E_{B_1}}\}\nonumber\\
& \quad \oplus N({_{\frac{\pi}{a}}A_0^+} \oplus {_{\frac{\pi}{a}}A_0^-} \oplus {_{\frac{\pi}{a}}A_1^+} \oplus {_{\frac{\pi}{a}}A_1^-}) \oplus \frac{N+1}{2}({_{\frac{\pi}{a}}B_1^+} \oplus {_{\frac{\pi}{a}}B_1^-}) \oplus \frac{N-1}{2}({_{\frac{\pi}{a}}B_0^+} \oplus {_{\frac{\pi}{a}}B_0^-})\\
\nonumber\\
\Gamma^{\mbox{\tiny phon}}_{\mbox{\tiny A, even}} = &
N( {_{0}A_{0}^{+}} \oplus {_{0}A_{0}^{-}} \oplus {_{0}A_{1}^{-}} \oplus {_{0}A_{1}^{+}} ) \oplus \frac{N}{2}( {_{0}B_{0}^{-}} \oplus {_{0}B_{0}^{+}} \oplus {_{0}B_{1}^{+}} \oplus {_{0}B_{1}^{-}} )\nonumber\\
& \quad \oplus N\sum_k(2{_{~~k}^{-k}E_{A_0}}\oplus 2{_{~~k}^{-k}E_{A_1}} \oplus {_{~~k}^{-k}E_{B_0}} \oplus {_{~~k}^{-k}E_{B_1}}) \oplus N(2{_{\frac{\pi}{a}}E_{A_0}^{A_1}} \oplus {_{\frac{\pi}{a}}E_{B_0}^{B_1}})\\
\nonumber\\
\Gamma^{\mbox{\tiny phon}}_{\mbox{\tiny Z, even}} = & 
N( {_{0}A_{0}^{+}} \oplus {_{0}A_{0}^{-}} \oplus {_{0}A_{1}^{-}} \oplus {_{0}A_{1}^{+}} \oplus {_{0}B_{0}^{+}} \oplus {_{0}B_{1}^{+}} ) 
 \oplus N\sum_k(2{_{~~k}^{-k}E_{A_0}}\oplus 2{_{~~k}^{-k}E_{A_1}} \oplus {_{~~k}^{-k}E_{B_0}}\oplus {_{~~k}^{-k}E_{B_1}})\nonumber\\
& \quad \oplus N({_{\frac{\pi}{a}}A_0^+}\oplus {_{\frac{\pi}{a}}A_0^-} \oplus {_{\frac{\pi}{a}}A_1^+}\oplus {_{\frac{\pi}{a}}A_1^-} \oplus {_{\frac{\pi}{a}}B_0^+}\oplus {_{\frac{\pi}{a}}B_1^+} )\\
\nonumber\\
\Gamma^{\mbox{\tiny phon}}_{\mbox{\tiny Z, odd}} = &
N( {_{0}A_{0}^{+}} \oplus {_{0}A_{0}^{-}} \oplus {_{0}A_{1}^{-}} \oplus {_{0}A_{1}^{+}} \oplus {_{0}B_{0}^{+}} \oplus {_{0}B_{1}^{+}} ) \oplus N\sum_k(2{_{~~k}^{-k}E_{A_0}}\oplus 2{_{~~k}^{-k}E_{A_1}} \oplus {_{~~k}^{-k}E_{B_0}}\oplus {_{~~k}^{-k}E_{B_1}})\nonumber\\
& \oplus N(2{_{\frac{\pi}{a}}E_{A_0}^{A_1}} \oplus {_{\frac{\pi}{a}}E_{B_0}^{B_1}})
\end{align}

\end{widetext}

\bibliographystyle{apsrev4-1}

\end{document}